# The electronic structure, crystal fields, and magnetic anisotropy in RECo$_5$ magnets

Zhen Zhang[1], Andrey Kutepov[2], Leonid Pourovskii[3,4], Vladimir Antropov[1,5]

[1]Department of Physics and Astronomy, Iowa State University, Ames, IA 50011, USA
[2]Department of Physics and Astronomy and Nebraska Center for Materials and Nanoscience, University of Nebraska-Lincoln, Lincoln, NE 68588, USA
[3] CPHT, CNRS, École polytechnique, Institut Polytechnique de Paris, 91120 Palaiseau, France
[4]Collège de France, Université PSL, 11 place Marcelin Berthelot, 75005 Paris, France
[5]Ames National Laboratory, U.S. Department of Energy, Ames, IA 50011, USA

The current progress in describing rare-earth-based magnets' electronic structure and magnetic properties is discussed. We use several currently popular electronic structure methods to show the typical values of critical parameters that define the physics of RECo$_5$ (RE = rare earth atom) materials. The magnetic moments and anisotropy of 4$f$ atoms are obtained using several approaches, including anisotropic 4$f$-charge density-constrained DFT and DFT+HI methods. We also suggest the introduction of 'penalty' functional for obtaining correct variational total energy in the traditional Hund's rule-constrained DFT-based techniques. The applicability and future extensions are discussed. The proposed combination of methods is potentially suitable for high-throughput computational searches of new rare-earth-containing magnetic materials.

## I. INTRODUCTION

The electronic structure of RE-based systems for years has been a very challenging topic for the computational physics of the solid state. While the qualitative theoretical understanding of such systems was formulated in the 30s[1] in the form of crystal field (CF) theory, the quantitative description still is not consistent. Modern theoretical description of the electronic structure is based on the density functional theory (DFT) that has been applied to RE-based systems since the 60s. With the development of the computational methods by the end of the 80s, it becomes clear that the DFT approach with its traditional local density approximation (LDA) fails to correctly describe positions of 4$f$ levels, their bandwidths, crystal field splittings and anisotropy of potential[2].

CF theory is a theory of the isolated atom surrounded by other atoms, and for RE atoms, it starts with the assumption that the electronic structure of 4$f$ electrons is like the structure of the isolated atom, modified only weakly by magnetic and electric interactions with the surrounding atoms. RE version of CF theory typically uses the following assumption about the strength of major interactions $E_{so} > E_{sp} \gg E_{cf}$[3]. In Fig. 1, we show the typical energies of such interactions for RE atoms in RECo$_5$ systems. Experimentally, such a ratio between these interactions is also well documented.

In describing a realistic electronic structure, one must consider the possible solid-state screening of Coulomb interaction. The LDA screening for 4$f$ states does not look correct, and 4$f$ states in such description appear at the Fermi level with relatively small 4$f$ spin splitting. This problem was corrected in the first applications of the LDA+$U$ method for Gd and CeSb[4–6]. Such ad-hoc addition of Hubbard $U$ term further splits 4$f$ levels away from the Fermi level and moves them into positions found by XPS or PES experiments[7]. This effect demonstrated that the weakening of hybridization of 4$f$ states with surroundings leads to the correct prediction of experimentally observed crystal and magnetic structures of 4$f$ systems[4–6]. The physical meaning of parameter $U$ in this scheme is not a 'true' Hubbard parameter as it mimics the combined action of two very large effects of LDA: uncontrolled weakness of Hubbard correlations and the presence of self-interaction errors. Further simplified calculations using an 'open core' approach (which corresponds to $U = \infty$) confirmed these conclusions[3].

Now, such LDA+$U$ and 'open core' methods[3,8,9], together with methods including self-interaction corrections[10], are being widely used in the crystal and magnetic structure predictions of RE systems. In general, all such treatments with $LS$ coupling and Hund's rule enforcement are in line with the assumption of the CF model, which is considered a 'standard model' of rare earths.

More recent developments of *ab initio* CF methodologies replace LDA 'open core' treatment of 4$f$ states by a Hubbard-I (HI) atomistic description in the framework of DFT+DMFT methods[11–13]. The 'open core' treatment of the 4$f$ electrons comes closest to the HI approximation since the LDA+$U$ approximation with the 'band' description of 4$f$ states is found to still strongly overestimate the hybridization and CF interactions resulting in the formation of dispersive energy states.

While the introduction of $U$-terms or 'open core' treatments allows to eliminate 'bad' participation of $f$-states in the hybridization, correcting the magnetic ground state and predicting reasonably well Curie or Neel temperatures, the 'secondary' magnetic properties, such as, for instance, magnetic anisotropy or magnetooptics cannot be completely corrected in this approach. Already, the initial application of LDA+$U$[4–6] demonstrated that 4$f$ states, while being more localized than in pure LDA, still do not have a structure corresponding to $LS$-coupling with Hund's rules (main assumptions of CF theory). Instead of an ideal atomic structure, all nearly localized one-electron $f$-states are strongly affected by CF of LDA. These CF shifts appear to be 10-30 times larger than experimentally known CF splittings. The calculations of the exchange field on the RE atom also produced numbers 4-5 times larger than experimentally estimated and used in CF models' applications[9]. LDA+$U$ weakly improves both one electron CF splitting and exchange fields[9], but the problem essentially remains. This is clearly seen in the calculations of magnetic anisotropy. Even if the LDA (or LDA+$U$) finds a state with the largest spin and orbital moments with the field along the z-direction, such Hund's state will be destroyed when the field is in-plane. Besides LDA CF splittings being overestimated, the orbital moment has a pure relativistic nature, and its value is not guaranteed in LDA. It means a corresponding to Hund's rule anisotropy of the 4$f$ density and crystal potential also is not guaranteed. In addition, when Hund's state is realized (for instance, m = -3 in CeSb[4-6]), the corresponding 4$f$ density and associated potential becomes anisotropic. Their interaction represents a self-interaction that destroys the proper energetics of 4$f$ levels. The suggestion to keep only the spherical average of the $f$-charge in the total charge density was made[6].

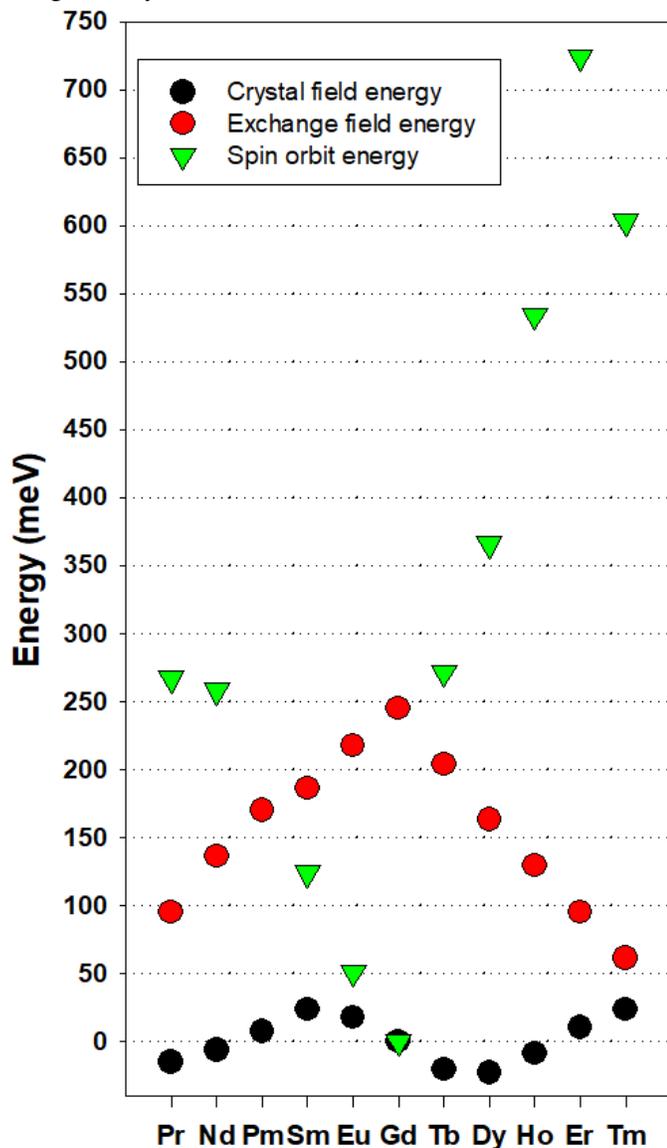

Fig. 1. Hierarchy of spin-orbit energy (green), exchange field (red), and crystal field energy (black) in RECo$_5$ magnets. Experimental lattice constants have been used.

Later several other ways to eliminate 'parasitic' self-interaction have been made in the different methods[14,15]. In such cases, the proper nearly atomistic order of levels ($LS$-coupling and Hund's rule) is retained[14,15].

Another opportunity to correct 'bad' anisotropic 4$f$ density is an 'orbital polarization' (OP) method[16,17], which is a particular case of more general LDA+$U$ construction with specific values of $U$ and $J$ parameters. In this case, only the Racah part of the Hartree-Fock matrix is retained. The inclusion of the Racah parameter ($E^3$ for $f$-states) then directly affects the anisotropic density and changes the OP accordingly. Correspondingly, this addition affects only CF related splittings. Unfortunately, LDA+OP practitioners, while calculating the magnetic anisotropy, did not analyze the resulting electronic structure and how the addition of $E^3$ affects anisotropic 4$f$ density. In all described above LDA+$U$ type of calculations, the total energy is not variational as the 'ad hoc' addition to the DFT energy is not compensated.

In 1997 Brooks et al.[18] suggested DFT constrained total energy calculations, which include described above elimination of self-interaction of the nonspherical part of the 4$f$ densities. The CF energies and the corresponding spin Hamiltonian parameters have been calculated from "first principles" demonstrating a good agreement with the experiment. Unfortunately, this method has not been further tested. However, the opportunity of having the correct total energy in the simplest approximation is still exciting.

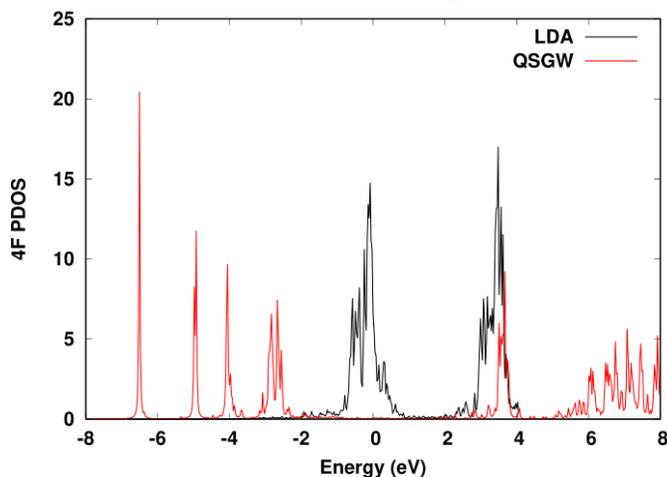

Fig. 2. Density of 4$f$-states of Sm atom in SmCo$_5$ obtained in LDA (black) and QSGW (red).

Among *ab initio* approaches, the QSGW method has been used for many materials as an alternative to DFT methods and is free of LDA self-interaction errors. In Fig. 2 we show the comparison of the results of LDA and QSGW methods for the density of states of Sm atom in SmCo$_5$. Clearly, the QSGW 4$f$-structure of occupied states is very different from LDA and is close to LDA+$U$ results. Our orbital decomposition indicates a very atomistic order of 4$f$ states in QSGW, which are very little affected by small CF splittings. Thus, the QSGW method can be considered as an *ab initio* alternative to study magnetic and optical effects induced by small CF splittings of RE atoms.

DFT+DMFT in HI approximation (DFT+HI)[19-21] combines many-body quasi-atomic treatment of the Hubbard interaction between localized 4$f$ electrons with DFT for the rest of the systems. It does not postulate *LS* coupling or Hund's rules and is able to identify deviation from them. DFT+HI appears rather successful in the description of structural, electronic, and magnetic properties of RE systems[12,13,22]. However, the straightforward DFT+HI also suffers from the DFT self-interaction problem. Ref.[11] thus combined DFT+HI with spherical averaging in the spirit of Brooks et al.[18]. Moreover, to effectively include the hybridization contribution to CF, it represented *f*-orbitals by extended Wannier functions formed from a narrow range of *f*-electron bands. Using this novel methodology, CF parameters for a range of RE materials, including RECo$_5$ systems[11,23], have been obtained in good agreement with the experimental estimates.

Unfortunately, due to many factors, the total energy calculations in DMFT-based methods currently cannot be performed with the accuracy suitable for the discussion of 'secondary' magnetic properties. Thus, due to the similarity to CF theory construction, one can use the results of CF theory to estimate properties of the 4*f* subsystem, while for the rest of the system, one can use DFT-related methods.

## II. METHODOLOGY

LDA calculations in Figs. 1 and 3 have been performed with the code FlapwMBPT[24,25]. This code implements a full-potential linearized augmented plane wave approach. A fully relativistic version of the code was used which is based on the Dirac equation. An open-core treatment of the 4*f* electrons with an exchange-correlation functional by Perdew and Wang[26] was implemented, and nonspherical components of the 4*f* charge density were ignored. To estimate the effect of the nonspherical components of the effective potential acting on the 4*f* orbitals, we also performed a series of calculations in the first proposed method where the effective potential was reduced to the spherical form (this was done only when evaluating the CF energies). The resulting state (Fig. 3) can be a mix of *LS*- and *jj*- couplings and can violate Hund's rule construction. QSGW calculations have been performed using the method described in Ref.[27].

Hund's rules constraining calculations have been performed using LMTO[4–6]. In our CF analysis, we employed the non-spin-polarized CF parameterization previously obtained for SmCo$_5$[23] using the DFT+HI approach of Ref.[11].

## III. HUND'S RULES CONSTRAINT (HRC) IN THE DENSITY FUNCTIONAL

The DFT methods with 'ad-hoc' terms added to the original DFT Hamiltonian are similar to noncollinear magnetic configuration calculations and do not, in general, correspond to an energy minimum. In such cases, it was suggested by Edwards[28] that one must impose some constraints with a minimization principle involving using Lagrange multipliers. In the second method considered in this paper, we are following the recipe of Ref.[29] and introduce a corresponding penalty functional to the traditional Hund's rules constraint (HRC)[10,30] on spin and orbital magnetic moment values and limit ourselves by considering only collinear states.

In this approach, the constrained total energy functional has

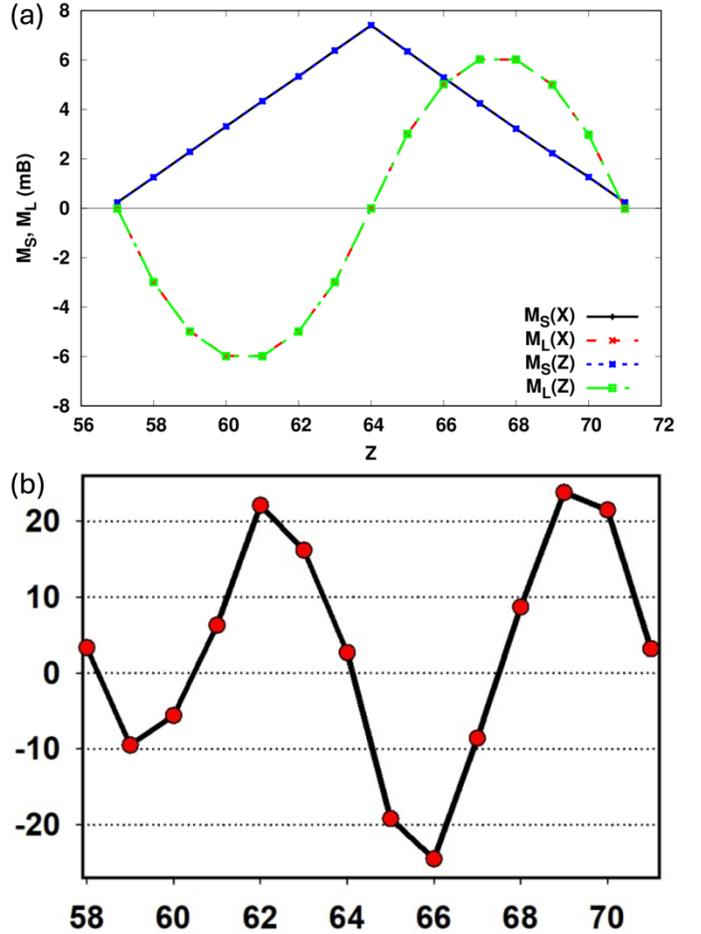

Fig. 3. (a) Magnetic spin ($M_S$) and orbital ($M_L$) moments on the rare earth atom in RECo$_5$ systems. X (Z) corresponds to the magnetic field in X (Z)-direction. (b) magnetic anisotropy of RE atoms (in meV) in RECo$_5$ obtained using the total energy in 4*f* anisotropic density constrained LDA 'open core' calculations.

the form

$$E = E_0 + E_p = E_0 + \sum_i \lambda_i \left( |\boldsymbol{M}_i| - \boldsymbol{e}_i \boldsymbol{M}_i \right)$$

where $E_0$ is the original DFT energy and $E_p$ is a traditional penalty functional. $\boldsymbol{M}_i$ is a total spin or orbital magnetic moment on site i. Both spin and orbital moments constraints are applied simultaneously to each 4*f* spin orbital only to reach the new magnetic state with imposed Hund's rule. Orbital moment constraint is like OP correction of Brooks[16] (see application for SmCo$_5$ in Ref.[31]), only now with the minimization principle enforced. Naturally, such constraining conditions can enforce not only *LS* or *jj* couplings but also *jl*, *Jj*, or any other non-uniform couplings with and without Hund's rule.

In this way, the results for spin and orbital moments on 4*f* states of the RE atom (Fig. 3) are obtained automatically, while the magnetism of all non-4*f* electrons in the systems is not constrained and obtained self-consistently.

Our results for magnetic anisotropy are shown in Fig. 4. Clearly, the results look similar to the total energy obtained in 4*f*-charge-constrained calculations. This is also a generic dependence of the CF energy (or CF magnetic anisotropy) known from CF theory. The variation of the sign of CF anisotropy follows oblate or prolate shapes of 4f electron cloud of tripositive RE ions in their ground states of the Hund's rule. Deviation at the beginning of the 4f row is related to the deviation from the Hund's rules and the formation of not localized states in CeCo$_5$. HRC for Ce systems should not be enforced. In the same way, the 'open core' or Hubbard I treatments cannot be applied to many systems containing Ce atoms.

The analysis of the contribution to the magnetic anisotropy in SmCo$_5$ in all constrained DFT calculations indicated a small contribution (<10%) from the spin-orbit energy. This is expected for Sm, while in 3d systems, the magnetic anisotropy nearly always follows the anisotropy of spin-orbit energy[32].

In the third considered method, we considered spherical averaging of 4f density in DFT+HI method and decomposed the Sm magnetic anisotropy[11] (Fig. 5). To that end, we employed the SmCo$_5$ CF and exchange-field parameters calculated by DFT+HI (Table VI of Ref.[23]) to set up the one-electron part of the single-site 4f Hamiltonian for a given direction of the exchange field (which is aligned to Co magnetization direction). The direction is parametrized by the azimuthal angle θ to the easy axis c within the (11$\bar{2}$0) hexagonal plane. Having diagonalized this Hamiltonian (with a rotationally invariant Hubbard interaction included) we computed the contribution of each term into the total energy from the ground-state 4f wavefunction.

The dominance of CF anisotropy $E_{CF}$ is evident from Fig. 5, while spin-orbit anisotropy $E_{SO}$ is about 15% of $E_{CF}$ and negative. Overall, the total influence of non-CF anisotropy terms appears at a level of 10%. Similar to the first method (Fig. 3), the powerful advantage of DFT-HI self-consistent calculations is the absence of any Hund's rules constraints so the resulting many body state can also be a mix of LS- and jj-couplings.

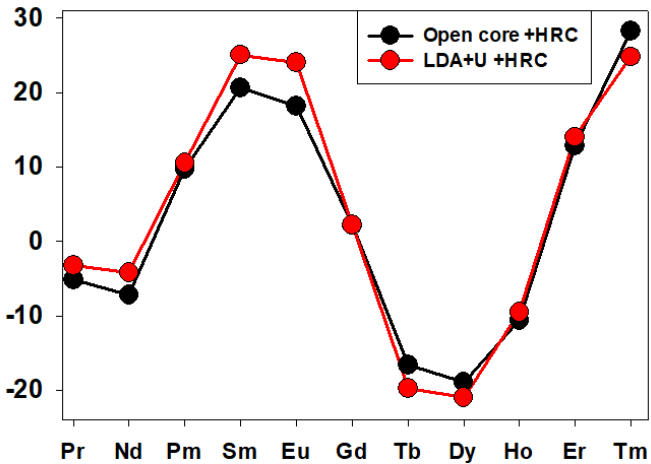

Fig. 4. The magnetic anisotropy of RECo$_5$ in the 'open core' LDA and LDA+$U$ ($U$ = 6.7 eV) with Hund's rules constraint enforced.

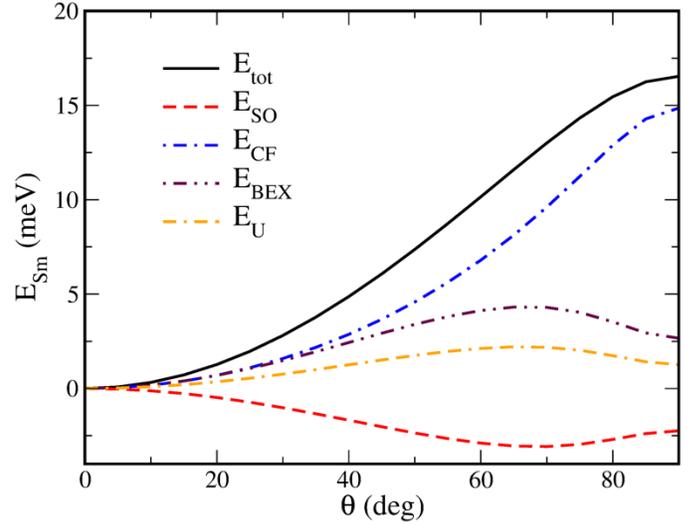

Fig. 5. The magnetic anisotropy energy of 4f states on Sm atom in SmCo$_5$ obtained from DFT+HI CF parameters as a function of the Co magnetization direction.

IV. CONCLUSION

The modern electronic structure computational methods are mature enough to produce reliable ground states and exited properties of such 4f-electron systems. However, such advances must be implemented in the standard electronic structure codes to avoid numerous difficulties of the past. Specifically, in DFT-based methods the correct description must include: (i) localized treatment of 4f electrons ('open core' or Hubbard I), (ii) introduction of Hubbard $U$ to reproduce the experimental spin splittings, (iii) the elimination of anisotropic 4f charge density to correct crystal field splittings. If the total energy is not accessibly, the crystal field theory with Wannier function implementation can be used. The approaches with the enforcement of Hund's rule, while also successful, have limited applicability and uncontrolled electronic structure. Non-DFT many body methods (QSGW) can also describe atomistic-like structure of 4f systems, however, due to absence of the reliable total energy should be also combined with crystal field theory to obtain required physical properties. Connection with the crystal field theory provides a much more complete picture of the exited states and the temperature effects. We emphasize that the above-proposed combination of these methods is, so far, the most flexible and accurate approach for material science progress in this area.

ACKNOWLEDGMENT

The work was primarily supported by the U.S. Department of Energy (DOE), Office of Basic Energy Sciences, Division of Materials Sciences and Engineering and performed at the Ames National Laboratory, which is operated for the U.S. DOE by Iowa State University under contract #DE-AC02-07CH11358. Z.Z. and A.K. acknowledge support by the U.S. DOE Established Program to Stimulate Competitive Research (EPSCoR) Grant No. DE-SC0024284. Computations were performed at the High-Performance Computing facility at Iowa

State University and the Holland Computing Center at the University of Nebraska.

Data Availability

The data that supports the findings of this study are available within the article.